# Beryllium-dihydrogen complexes on nanostructures

Hoonkyung Lee,[1,*] Bing Huang,[2] Wenhui Duan,[2] and Jisoon Ihm[1]

[1]Department of Physics and Astronomy, Seoul National University, Seoul, 151-747, Korea

[2]Department of Physics, Tsinghua University, Beijing 100084, People's Republic of China

ABSTRACT

Using the pseudopotential density functional method, we find that a Be atom on a nanostructure with $H_2$ molecules forms a Be-dihydrogen complex through the hybridization of the Be $s$ or $p$ orbits with the $H_2$ σ orbits and the binding energy of the $H_2$ molecules is in the range of ~0.3−0.8 eV/$H_2$. We also study Be-dihydrogen complexes on various nanostructures and demonstrate the feasibility of the application of the complexes to a hydrogen storage medium that operates near room temperature and ambient pressure.

Transition metal (TM)-dihydrogen complexes[1] by the hybridization of the TM $d$ orbits with the $H_2$ $\sigma$ or $\sigma^*$ orbits (i.e., the so-called Kubas interaction) have been of interest for the hydrogen storage purposes because of the potential of the application to a hydrogen storage material that operates at room temperature and ambient pressure. Theoretical studies have shown that TM (Sc, Ti, V, and Ni)-decorated nanomaterials adsorb $H_2$ molecules with a desirable binding energy of ~0.2−0.6 eV by the Kubas interaction.[2-6] The hydrogen storage capacity reaches the gravimetric goal 9 wt% of the Department of Energy (DOE) by the year 2015.[7] Recently, it has been found that calcium-decorated nanostructured materials adsorb $H_2$ molecules as much as 5 wt% with a binding energy of ~0.2 eV/$H_2$ by the hybridization of the empty Ca $3d$ states with the $H_2$ $\sigma$ states.[8-12] This binding mechanism of $H_2$ molecules on Ca atom is similar to that on TM atoms.

Here, fundamental questions arise: Do systems including $d$ orbits only induce a dihydrogen complex? Can systems only including $s$ and $p$ orbits cause a dihydrogen complex? In this paper, we show that, similarly to TM- or Ca-dihydrogen complexes, $H_2$ molecules bind to beryllium (Be)-decorated nanomaterials with a binding energy of ~0.3−0.8 eV in a form of dihydrogen through the hybridization of the Be $s$ or $p$ orbits with the $H_2$ $\sigma$ orbits. Using the equilibrium grand partition function, we find that $H_2$ molecules bind to the Be atom at 25 $^o$C and 30 atm and they are released at 100 $^o$C and 30 atm. This result exhibits that Be-dihydrogen complexes may be applicable to a hydrogen storage medium that operates near room temperature and ambient pressure.

All our calculations were carried out using the pseudopotential density functional method with the plane-wave-based total energy minimization[13]. The exchange correlation energy functional of generalized gradient approximation (GGA)[14] was used, and the kinetic energy cutoff was taken to be 400 eV. The optimized atomic positions were relaxed until the Hellmann-Feynman force on each atom meets less than 0.01 eV/Å. The supercell calculations throughout were employed where the adjacent structures are separated by over 10 Å.

We choose benzene ($C_6H_6$), pentane ($C_5H_{12}$), aniline ($C_6H_5NH_2$), and ethane-1,2-diol ($C_2O_2H_6$) as representatives for $sp^2$-bonded carbon, $sp^3$-bonded carbon, amine (-$NH_2$) group, and hydroxyl (-OH) group, respectively. We replace one hydrogen atom of those molecules with a single Be atom to make complexes consisting of Be atoms and organic molecules. From now on, we denote the above four complexes as Be-$sp^2$-carbon, Be-$sp^3$-carbon, Be-NH, and Be-O, respectively. The binding energies of the Be atom on $sp^2$-carbon, $sp^3$-carbon, -$NH_2$ group, and -OH group are calculated to be 2.7, 1.9, 3.0, and 4.3 eV, respectively. We find that the Be atom with up to two $H_2$ molecules forms a Be-dihydrogen complex as shown in Figs. 1(a)−1(c) except for the Be-O complex which does not form any complexes with $H_2$ molecules. The bond length of the $H_2$ molecules is substantially elongated to ~0.86 Å from 0.75 Å of isolated molecule and the distance between the Be atom and the $H_2$ molecule is 1.60 Å. The calculated binding energy of the $H_2$ molecules on Be-$sp^2$-carbon, Be-$sp^3$-carbon, and Be-NH as a function of adsorbed $H_2$ molecules is presented in Table I.

We also take into account boron (B; valence electronic configuration: $2s^22p^1$) and magnesium (Mg; $3s^2$) atom instead of Be ($2s^2$) atom to examine whether $H_2$ molecules are adsorbed as in the case of the Be atom. We find that, unlike the case of Be-decorated organic molecules, $H_2$ molecule binds dissociatively to the B atom with a binding energy of ~1 eV. In contrast, $H_2$ molecules do not bind to Mg-decorated organic molecules. On the other hand, according to a recent paper,[15] lithium (Li; $2s^1$)-decorated organic molecules adsorb up to six $H_2$ molecules per Li atom with a binding energy of ~0.1 eV by the polarization of $H_2$ molecules. Therefore, the adsorption of $H_2$ molecules on Be atom is in sharp contrast to $H_2$ binding to Mg, B, and Li atoms.

Now, we consider pristine and boron-doped fullerenes for Be decoration to examine the dependence of $H_2$ binding as backbone materials. For a pristine fullerene ($C_{60}$), we find that a Be atom prefers to be adsorbed on the C-C bond consisting of two hexagons and pentagons. The calculated binding energy of the Be atom is 0.7 eV and the distance between the carbon atom and the Be atom is 1.63 Å. As in the case of Be-organic complexes, up to two $H_2$ molecules bind to the Be atom with the binding energy of 0.83 and 0.79 eV/$H_2$ for 1 and 2 $H_2$ molecules, respectively (Figure 2(a)). The distance between the Be atom and the $H_2$ molecule is ~1.6 Å and the bond length of the $H_2$ molecule is 0.87 Å. In contrast, for a B-doped fullerene ($C_{48}B_{12}$), a Be atom prefers to be attached to the hexagon center including two B atoms. The calculated binding energy of the Be atom is considerably increased up to 3.2 eV compared to the value of 0.7 eV in the Be-decorated $C_{60}$ and the distance between the Be atom and the B atom is 1.82 Å. The number and the binding energy of the $H_2$ molecules is reduced to 1 and 0.42 eV, respectively, compared to 2 and 0.83 eV in the Be-decorated $C_{60}$. The distance between the Be atom and the $H_2$ molecule is ~1.6 Å and the bond length of the $H_2$ molecule is 0.79 Å. We also consider other carbon-based materials: a B-doped CNT, a di-vacancy CNT, and a zigzag graphene nanoribbon (ZGNR). We find that Be atoms prefer to be adsorbed on the B or defect sites on CNTs and the edge of the ZGNR with the binding energy of 1.70, 1.03, and 0.82 eV per Be atom, respectively. Up to one, two, and one $H_2$ molecules bind to each Be atom on the B-doped CNT, di-vacancy CNT, and ZGNR, respectively, as shown in Figs. 2(c)–2(e). Therefore, the number of adsorbed $H_2$ molecules depends on the backbone structures.

We next investigate the origin of the difference of the number of adsorbed $H_2$ molecules as backbone materials. Figure 3 exhibits the projected or partial density of states for the Be-decorated $C_{60}$ and $C_{48}B_{12}$ before and after the adsorption of $H_2$ molecules. For the Be-decorated $C_{60}$, we find that the unoccupied Be $p_x$, $p_z$, or $s$ states are hybridized with the $H_2$ states when two $H_2$ molecules bind to the Be atom as described in Figs. 3(a)−3(c). In contrast, for the Be-decorated $C_{48}B_{12}$, the Be $p_x$ state does not contribute to the $H_2$ binding to the Be atom because the Be $p_x$ state is hybridized with the $C_{48}B_{12}$ states as shown in Fig. 3(e). Therefore, the unoccupied Be $p_x$ states play a crucial role of binding up to 2 $H_2$ molecules in our systems, explaining the different number of adsorbed $H_2$ molecules as Be-decorated structures. On the other hand, it has been found that Be ion (i.e., $Be_2^{2+}$) with $H_2$ molecules forms Be ion-dihydrogen complexes, $Be_2H_4^{2+}$ in a vacuum.[16] The Be-H and H-H distances in the $Be_2H_4^{2+}$ complex are 1.62 and 0.80 Å, respectively, which are consistent with our results. Unlike Be ion, a Be atom with $H_2$ molecules in a vacuum does not make any complexes. We believe that $H_2$ molecules adsorb on the Be atom on nanostructures in our systems since the Be atom acts like Be ion by charge transfer from the Be *s* states to the nanostructures as exhibited in Figs. 3(b) and 3(e).

To investigate the question of whether Be-dihydrogen complexes might be employed for hydrogen storage as in case of TM-dihydrogen complexes[9], we explore the thermodynamics for the adsorption of $H_2$ molecules on a Be atom as a function of the pressure and the temperature. For equilibrium conditions between the hydrogen gas and the adsorbed $H_2$ molecules, the (fractional) occupation number *f* per site is as follows:

$$f = \frac{\sum_{n=0}^{N_{max}} g_n n e^{n(\mu - \varepsilon_n)/kT}}{\sum_{n=0}^{N_{max}} g_n e^{n(\mu - \varepsilon_n)/kT}} \quad (1)$$

where the maximum number of adsorbed $H_2$'s per site is $N_{max}$, $g_n$ is the degeneracy of the configuration for a given *n*, $\mu$ is the chemical potentials of $H_2$ gas, $-\varepsilon_n$ (>0) is the binding energy of $H_2$ molecules per $H_2$ at a given *n*, and *k* and *T* are the Boltzmann constant and the temperature, respectively. In the calculation of *f*, the experimental chemical potentials for $H_2$ gas is employed, and the binding energy of the $H_2$ molecules is subtracted by 35% from the calculated (static) binding energy in Table I because of the zero-point vibration energy.[4] Figure 4 shows the occupation-pressure-temperature (*f-p-*T) diagram for the Be-$sp^2$-carbon and Be-$sp^3$-carbon. The calculated usable number of $H_2$ molecules per Be atom at ambient conditions (*f* at the adsorption conditions of 30 atm and 25 °C minus *f* at the desorption conditions of 3 atm and 100 °C) for all the structures is presented in Table II. For the Be-decorated $C_{60}$, the usable number of $H_2$ molecules per Be atom is zero because two adsorbed $H_2$ molecules remain not released on the Be atom at the desorption conditions. This is attributed to the Gibbs factor for the adsorption of 2 $H_2$ molecules that dominates ($e^{2(\mu-\varepsilon_2)/kT} \gg e^{(\mu-\varepsilon_1)/kT} \gg 1$ where $\varepsilon_1$ and $\varepsilon_2$ are –0.54 and –0.51 eV, respectively.) at the adsorption ($\mu = -0.23$ eV) and desorption ($\mu = -0.39$ eV) conditions. In contrast, the usable number of H$_2$ molecules on the Be-$sp^2$-carbon is 1.98 because the Gibbs factor for the adsorption for 2 H$_2$ molecules dominates ($e^{2(\mu-\varepsilon_2)/kT} >> 1 >> e^{(\mu-\varepsilon_1)/kT}$, $\varepsilon_1$ and $\varepsilon_2$ are $-0.07$ and $-0.25$ eV) at the adsorption conditions and the term becomes negligible ($e^{2(\mu-\varepsilon_2)/kT} << 1$) at the desorption conditions. We find that H$_2$ molecules adsorbed on the Be atom in the Be-$sp^2$-carbon and Be-$sp^3$-carbon are almost fully usable at ambient conditions. These results show that Be-dihydrogen complexes on nanostructures might be used for a hydrogen storage medium which operates at room temperature and ambient pressure.

In conclusion, using the pseudopotential density functional method, we have demonstrated that beryllium atom with H$_2$ molecules on nanostructures forms Be-dihydrogen complexes, similarly to transition metal-dihydrogen complexes. Because of a desirable binding energy of the dihydrogen (~0.4 eV/H$_2$), Be-decorated nanostructures may be suitable as a hydrogen storage medium that operates at room temperature and ambient pressure.

**ACKNOWLEDGMENT.** This research was supported by the Center for Nanotubes and Nanostructured Composites funded by the Korean Government MOST/KOSEF, and the Korean Government MOEHRD, Basic Research Fund No. KRF-2006-341-C000015. The work at Tsinghua was supported by the Ministry of Science and Technology of China, the Natural Science Foundation of China and the Ministry of Education of China.

*E-mail: hkiee3@snu.ac.kr

TABLE I. Calculated binding energy of a Be atom (eV) on a $sp^2$-carbon, $sp^3$-carbon, amine group, $C_{60}$, and $C_{48}B_{12}$, and the binding energy of $H_2$ molecules (eV/$H_2$) on the Be atom as a function of the number of adsorbed $H_2$ molecules.

| Materials | Be | 1 $H_2$ | 2 $H_2$ |
|---|---|---|---|
| **Be-$sp^2$-carbon** | 2.7 | 0.22 | 0.39 |
| **Be-$sp^3$-carbon** | 1.9 | 0.11 | 0.33 |
| **Be-NH** | 3.0 | 0.76 | 0.35 |
| **Be-$C_{60}$** | 0.7 | 0.83 | 0.79 |
| **Be-$C_{48}B_{12}$** | 3.2 | 0.42 | |

TABLE II. Comparison for the usable number of $H_2$ molecules in Be-decorated nanomaterials. Here $f \equiv N_{ads}$ is the number of adsorbed $H_2$'s per Be atom at the condition of adsorption (30 atm-25 °C), and $f \equiv N_{des}$ is the number of adsorbed $H_2$'s per Be atom at the condition of desorption (3 atm-100 °C). $N_{max}$ is the number of attached number of $H_2$ molecules. The usable number of $H_2$ molecules per Be atom is obtained from $N_{use} = N_{ads} - N_{des}$.

| Materials | $N_{max}$ | $N_{ads}$ | $N_{des}$ | $N_{use}$ |
|---|---|---|---|---|
| **Be-$sp^2$-carbon** | 2 | 1.99 | 0.01 | 1.98 |
| **Be-$sp^3$-carbon** | 2 | 1.63 | 0.00 | 1.63 |
| **Be-NH** | 2 | 1.00 | 1.00 | 0.00 |
| **Be-$C_{60}$** | 2 | 2.00 | 2.00 | 0.00 |
| **Be-$C_{48}B_{12}$** | 1 | 0.97 | 0.10 | 0.87 |

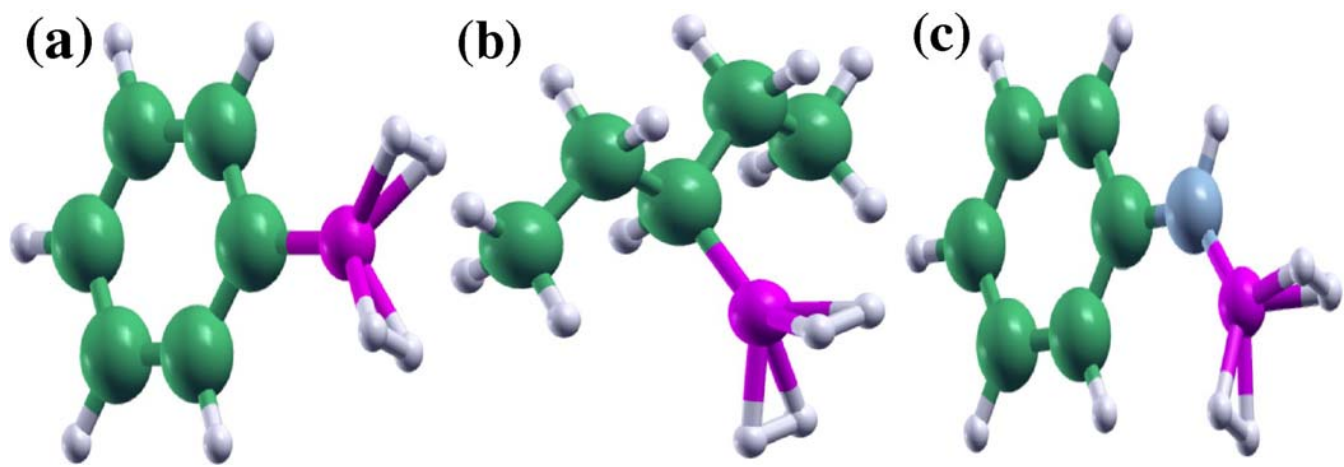

FIG. 1. (Color online) Optimized atomic geometries of Be-decorated organic molecules with maximally adsorbed $H_2$ molecules. (a) Be-decorated benzene with two $H_2$ attached, (b) Be-decorated pentane with two $H_2$ attached, and (c) Be-decorated aniline with two $H_2$ attached. Green, pink, white, and grey dots indicate carbon atom, beryllium atom, hydrogen atom, and nitrogen atom.

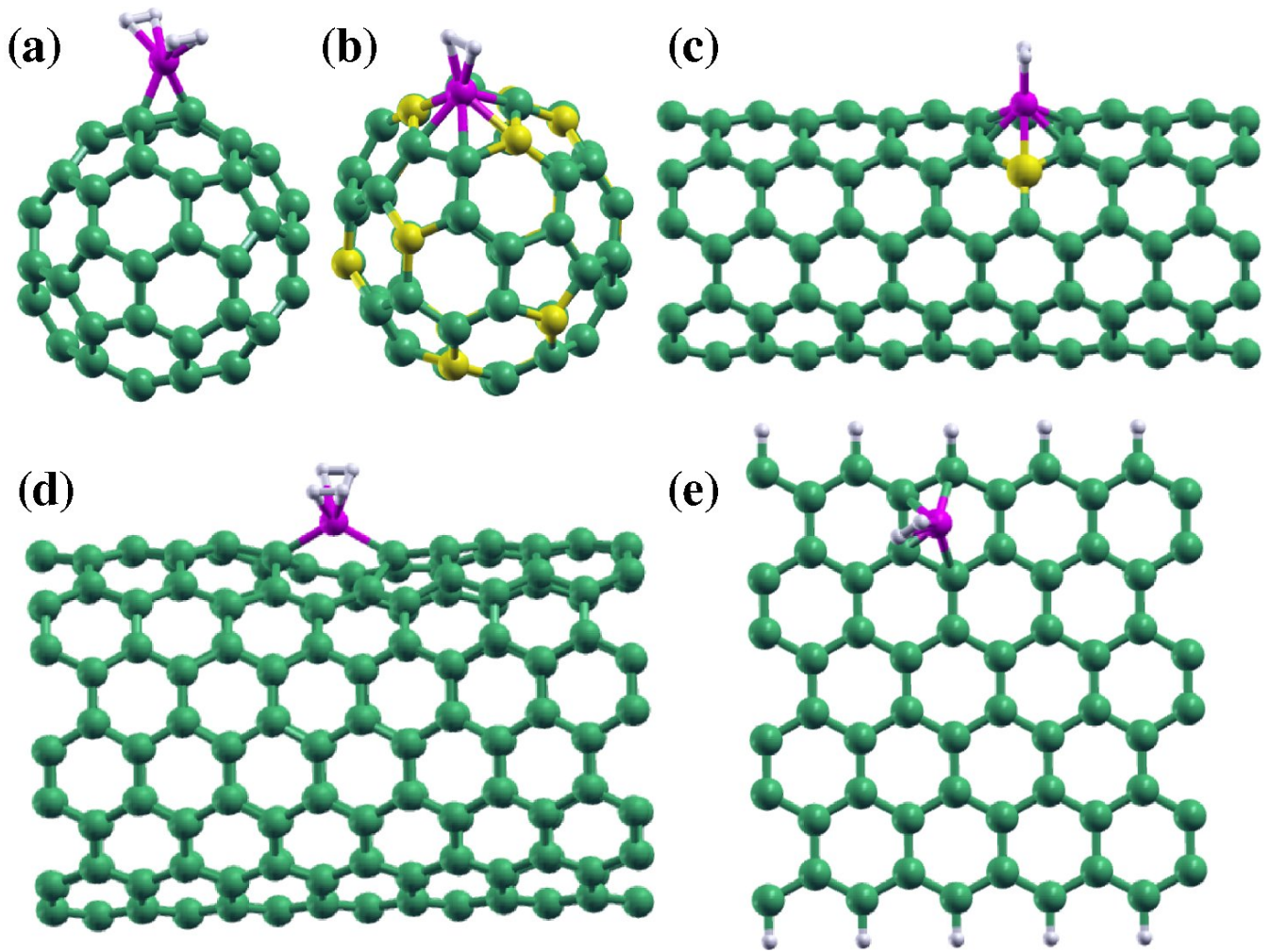

FIG. 2. (Color online) Optimized atomic geometries for a Be-decorated (a) $C_{60}$ with two $H_2$ attached, (b) $C_{48}B_{12}$ with one $H_2$ attached, (c) B-doped CNT with one $H_2$ attached, (d) di-vacancy CNT with two $H_2$ attached, and (e) ZGNR with one $H_2$ attached. Yellow dots indicate boron atoms.

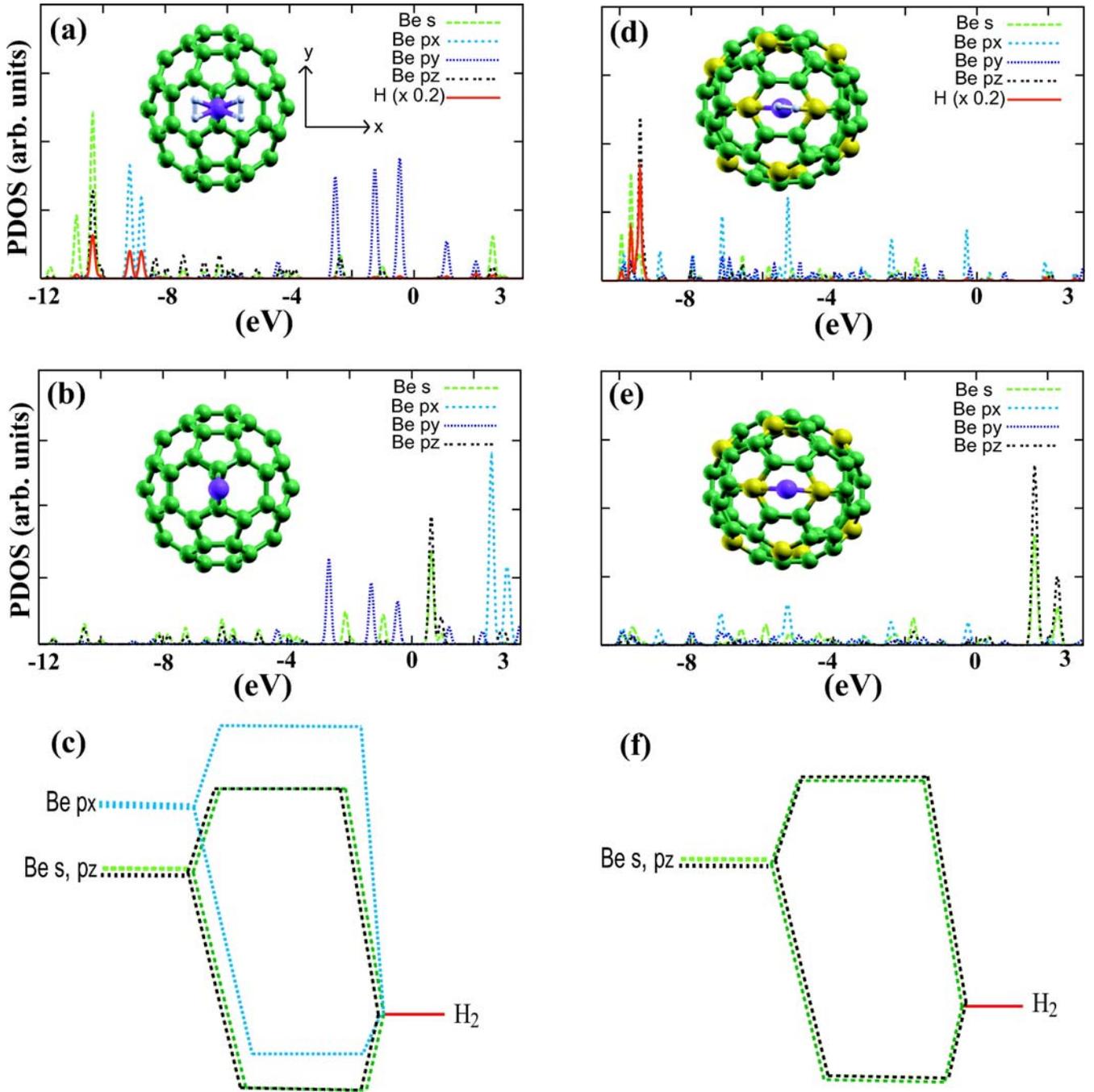

FIG. 3. (Color online) PDOS of the Be atom and $H_2$ molecules for (a) Be-decorated $C_{60}$ when two $H_2$ molecules adsorb, (b) Be-decorated $C_{60}$, (d) Be-decorated $C_{48}B_{12}$ when one $H_2$ molecule adsorbs, and (e) Be-decorated $C_{48}B_{12}$. The Fermi level is set to zero. (c) and (f) Schematic picture of the hybridization of the Be $p$ and $s$ states with $H_2$ states in Be-decorated $C_{60}$ and $C_{48}B_{12}$, respectively.

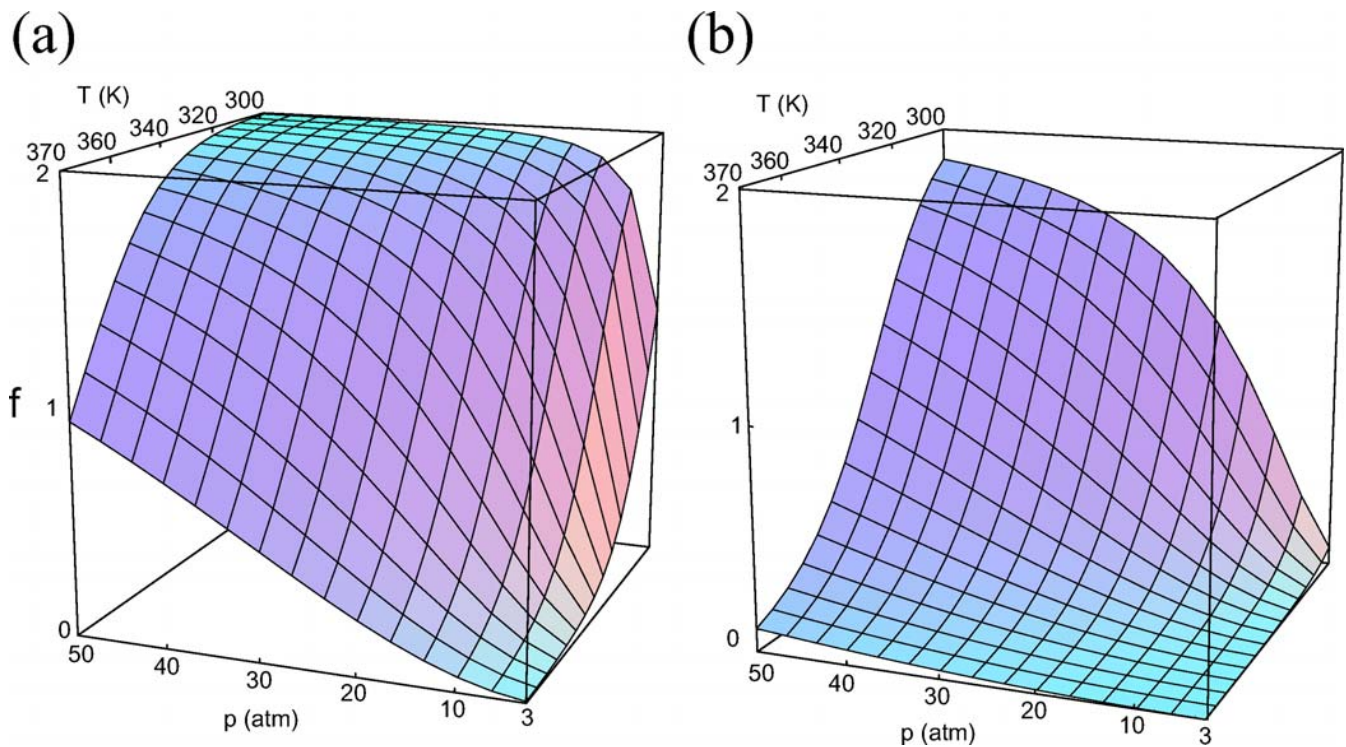

FIG. 4. (Color online) Occupation number of $H_2$ molecules as a function of the pressure and the temperature ($f$ - $p$ - $T$ diagram) on (a) Be-$sp^2$-carbon and (b) Be-$sp^3$-carbon.